\documentstyle[aaspp4]{article}

\newcommand\kms{\ifmmode{\;{\rm km}\;{\rm s}^{-1}}
 \else{~${\rm km}\;{\rm s}^{-1}$}\fi}
\newcommand\mass{\ifmmode{\cal M}\else{${\cal M}$}\fi}

\begin{document}

\title{The Baryonic Tully-Fisher Relation}
\author{S.S. McGaugh\altaffilmark{1},
J.M. Schombert\altaffilmark{2},
G.D. Bothun\altaffilmark{2},
and W.J.G. de Blok\altaffilmark{3,4}}
\altaffiltext{1}{Department of Astronomy, University of Maryland,
        College Park, MD 20742-2421; ssm@astro.umd.edu}
\altaffiltext{2}{Department of Physics, University of Oregon, Eugene, OR
97403;
js@abyss.uoregon.edu; nuts@bigmoo.uoregon.edu}
\altaffiltext{3}{Australia Telescope National Facility, PO Box 76,
                   Epping, NSW 1710, Australia; edeblok@atnf.csiro.au}
\altaffiltext{4}{Bolton Fellow}

\begin{abstract}
We explore the Tully-Fisher relation over five decades in stellar mass in
galaxies with circular velocities ranging over
$30 \lesssim V_c \lesssim 300\kms$.
We find a clear break in the optical Tully-Fisher relation:  field
galaxies with $V_c \lesssim 90\kms$ fall below the relation
defined by brighter galaxies.  These faint galaxies are however very gas
rich; adding in the gas mass and plotting baryonic disk mass
$\mass_d = \mass_* + \mass_{gas}$ in place of luminosity
restores a single linear relation.
The Tully-Fisher relation thus appears fundamentally to be
a relation between rotation velocity and total baryonic mass
of the form $\mass_d \propto V_c^4$.
\end{abstract}


\section{Introduction}

The relation between luminosity and rotation velocity for galaxies is well
known (Tully \& Fisher 1977\markcite{TF}).  It has been used extensively
in estimating extragalactic distances (e.g.,
Sakai et al.\ 2000\markcite{HSTkey}, Tully \& Pierce
2000\markcite{TP}), and it provides a critical constraint
on galaxy formation theory (Dalcanton, Spergel, \& Summers
1997\markcite{D97}; McGaugh \& de Blok 1998\markcite{MdB98a};
Mo, Mao, \& White 1998\markcite{MMW};
Steinmetz \& Navarro 1999\markcite{SN}; van den Bosch 1999\markcite{vdB}).
However, the physical basis of the Tully-Fisher relation remains unclear.

The requirements of the empirical Tully-Fisher relation are simple, but
the steep slope and small scatter are difficult to understand.
Luminosity must trace total (dark plus luminous) mass,
which in turn scales exactly with circular velocity.
Considerable fine-tuning is required to obtain these strict
proportionalities (McGaugh \& de Blok 1998\markcite{MdB98a}).
The intrinsic properties of dark halos are not
expected to be as tightly correlated as observed (Eisenstein \& Loeb
1995\markcite{EL}).  The mapping from the properties of dark matter
halos to observable quantities
should introduce more scatter, not less.
Somehow the baryons ``know'' precisely how many stars to form.

Let us suppose that, for whatever fundamental reason, there does exist a
universal relationship between total mass and rotation velocity of the
form $\mass_{tot} \propto V_c^b$.  The empirical Tully-Fisher relation then
follows if luminosity traces mass:
\begin{equation}
L = \Upsilon_*^{-1} f_* f_d f_b \mass_{tot},
\end {equation}
where $f_b$ is the baryon fraction of the universe,
$f_d$ is the fraction of the baryons associated with a particular galaxy
halo which reside in the disk, $f_*$ is the fraction of disk baryons in the
form of stars, and $\Upsilon_*$ is the mass-to-light ratio of the stars. 
Each of the pieces which intervene between $L$ and
$\mass_{tot}$ must be a nearly universal constant shared by all disks in
order to maintain the strict proportionality the Tully-Fisher relation
requires.  Cast in this form, 
the traditional luminosity-linewidth relation is a sub-set of
a more fundamental relation between {\it baryonic mass\/} and
{\it rotational velocity}.   In this context, one would expect
to find galaxies which
deviate from the luminosity-linewidth relation because much of their
baryonic mass is not in the form of stars.  For example, a gas rich
galaxy should appear underluminous
for its circular velocity, but would, after
correction for the gas content, fall on the underlying
``Baryonic Tully-Fisher relation''
(cf.\ Freeman 1999\markcite{Ken}).

In this paper, we specifically test this premise by constructing the
luminosity-linewidth and Baryonic Tully-Fisher relations for a sample of
late type galaxies that span a much
larger range of luminosities than any previously available sample.
Section 2 describes the data we employ.  Section 3 discusses the results
and \S 4 explores some of their implications.  A summary is given in \S 5.
All distance dependent quantities assume $H_0 = 75\kms\;{\rm Mpc}^{-
1}$.

\section{Data}

We employ several data sets to maximize the dynamic range over which
we
can explore the Tully-Fisher relation.  The different data sets have
photometry in different pass bands.  To put the data on the same system,
and get at the question
of the underlying mass, we assume a stellar mass-to-light ratio for each
pass
band.  Stellar mass is most directly traced by the redder pass bands, so
we
adopt these when possible.  

For galaxies with $V_c \gtrsim 100\kms$ we use
the extensive $H$-band data for late type cluster spirals of Bothun et al.\
(1985\markcite{cluster}).  Circular velocities are estimated as half of the
linewidth $W_{20}$.  For galaxies of lower rotation velocity,
we use the data for late type dwarf low surface brightness galaxies
from the survey of Schombert, Pildis, \& Eder (1997\markcite{Dgals}).  
This is currently the largest sample of field dwarf galaxies with both
linewidths $W_{20}$ and HI masses (Eder \& Schombert 1999\markcite{ES})
and red band photometry (Pildis, Schombert, \& Eder 1997\markcite{PSE}).
The photometry provides $I$-band magnitudes and axial ratios for
inclination estimates.  Of these galaxies, only those with axial
ratios $b/a < 0.71$, corresponding to
$i > 45^{\arcdeg}$ for an intrinsic axial ratio of $q_0 = 0.15$, are used in
order to minimize $\sin(i)$ errors.  These nevertheless contributes
substantially to the scatter, as inclinations estimated from the axial
ratios of dim galaxies are intrinsically uncertain.
The data for these faint galaxies extend the Tully-Fisher relation to much
lower luminosities and circular velocities than have been explored
previously.

The fundamental rotation velocity of interest here is the flat portion of
the rotation curve, $V_{flat}$.  Presumably, the line width $W_{20}$
commonly employed in Tully-Fisher work is an adequate 
indicator of $V_{flat}$.  As a check on this, we also employ the data
of Verheijen (1997\markcite{Marc}) and McGaugh \& de Blok
(1998\markcite{MdB98a})
for which $V_{flat}$ is measured from resolved rotation curves.
The data of Verheijen (1997\markcite{Marc}) are $K'$-band
data for spiral galaxies in the UMa cluster (Tully et al.\
1996\markcite{TVPHW}), while the data discussed by McGaugh \& de Blok
(1998\markcite{MdB98a}) are $B$-band data drawn from a variety of sources.

The two red-band data sets, the $H$-band data of Bothun et al.\
(1985\markcite{cluster}) at the bright end, and the $I$-band dwarf galaxy
sample at the faint end, together suffice to define a Tully-Fisher relation
over five decades in stellar mass.  The rotation curve samples
are consistent with these data.
For comparison, we also examine the gas rich, late type galaxy sample of
Matthews, van Driel, \& Gallagher (1998\markcite{MvDG}).
Their $B$-band data are entirely consistent with our own data, provided
we make the same inclination
cut, $i > 45^{\arcdeg}$.  Though this inclination limit is of obvious
importance, it is interesting to note that including or excluding the
galaxies they note as
having strongly asymmetric or single-horned HI profiles makes no
difference to the result.

In all cases, we have simply taken the data as given by each source.
Aside from the necessary inclination correction, we have not made any
corrections for internal extinction or for
non-circular motions (shown to be small for late type systems by Rix \&
Zaritsky 1995\markcite{RZ} and by Beauvais \& Bothun
1999\markcite{BB}).  That the data treated in this way produce a good
Tully-Fisher relation indicates, to first order,
that these effects are not important.

\section{Results}

Figure 1 illustrates the Tully-Fisher relation for the combined data sets.
Two versions are shown:  in (a) the stellar mass is plotted in place of
luminosity, and in (b) the total luminous baryonic mass is shown.  In order
to place the data sets using different band passes for photometry on the
same scale, we convert luminosity to stellar mass assuming a fixed
mass-to-light ratio $\Upsilon_*$ for each band.  The value of the
mass-to-light ratio appropriate to the stellar populations of
late type galaxies with ongoing star formation has been examined in
detail by de Jong (1996\markcite{dJ4}).  We adopt his
model for a 12 Gyr old, solar metallicity population with a constant
star formation rate and Salpeter IMF.  The adopted mass-to-light ratios
are $\Upsilon_*^B = 1.4$, $\Upsilon_*^I = 1.7$, $\Upsilon_*^H = 1.0$,
and\footnote{For the mean $H-K'$ color of late type galaxies given by
de Jong (1996\markcite{dJ4}), $\Upsilon_*^H = 1.2 \Upsilon_*^{K'}$.}
$\Upsilon_*^{K'} = 0.8\; \mass_{\sun}/L_{\sun}$.
These $K'$ and $I$-band mass-to-light ratios are
consistent with the maximum disk fits to the bright galaxies of Verheijen
(1997\markcite{Marc}) and Palunas (1996\markcite{PP}).
We do of course expect variation in stellar populations and their
mass-to-light ratios.  
This should be modest in the redder bands, especially $H$ and $K'$,
which are not very sensitive to differences in star formation history.
The $I$-band mass-to-light
ratio is not very sensitive to metallicity (Worthey 1994\markcite{Guy}),
so this should suffice for the fainter galaxies which in any case are
dominated by gas mass.  The $B$-band is a less robust indicator
of stellar mass, so we do not include these data in the fit in Figure 1(b).
While the absolute normalization of stellar mass-to-light ratios remains
uncertain, tweaking the adopted values has no effect on the basic result.

The stellar mass plotted in Figure 1(a) is simply $\mass_* = \Upsilon_*
L$, so this plot is directly analogous to the conventional
luminosity-linewidth diagram.  The baryonic disk mass plotted
in Figure 1(b) is the sum of stars and gas,
$\mass_d = \mass_* + \mass_{gas}$.  The mass in gas is taken from the
observed HI mass with the standard correction for helium and metals:
$\mass_{gas} = 1.4 \mass_{HI}$.
It appears that molecular gas is not a significant mass component in these
late type galaxies (Schombert et al.\ 1990\markcite{SCO};
de Blok \& van der Hulst 1998\markcite{BCO};
Mihos, Spaans, \& McGaugh 1999\markcite{MSM};
Gerritsen \& de Blok 1999\markcite{GdB}).

There have long been hints (e.g., Romanishin, Strom, \& Strom
1983\markcite{RSS}) that faint galaxies fall below the extrapolated
Tully-Fisher relation for bright galaxies.  Matthews et al.\
(1998\markcite{MvDG}) and Stil \& Israel (1999\markcite{SI})
claim to see this in their samples.  However,
it is not clear from their data in Figure 1.  The apparent discrepancy in our
results stems not from a difference in the data, but from what is taken to
define the Tully-Fisher relation.  Matthews et al.\ (1998\markcite{MvDG})
and Stil \& Israel (1999\markcite{SI}) compare their data to lines fit to
the $B$-band data of brighter galaxies.  These fiducial lines 
have a shallow slope which considerably overpredicts the luminosities of
faint galaxies when extrapolated to low circular velocity.  It is not clear
that it is safe to extrapolate the slope in this fashion.  Extinction 
appears to be relatively more important in brighter
galaxies, with careful corrections giving steeper slopes
(Tully et al.\ 1998\markcite{extinct}).
Samples of galaxies with low intrinsic extinctions also
give considerably steeper
$B$-band slopes (Sprayberry et al.\ 1995\markcite{lsbtf};
Verheijen 1997\markcite{Marc};
McGaugh \& de Blok 1998\markcite{MdB98a}).  The $H$-band
data of Bothun et al.\ (1985\markcite{cluster}) and the $K'$-band data of
Verheijen (1997\markcite{Marc}), two bands where extinction is minimal,
also indicate steep slopes.  A steep slope is also supported by the
calibration of the Tully-Fisher relation from the HST Key Project
(Sakai et al.\ 2000\markcite{HSTkey}).  Such a slope
eliminates the discrepancy reported by Matthews et al.\
(1998\markcite{MvDG}) and by Stil \& Israel (1999\markcite{SI}).

Nevertheless, it is now clear from the larger dwarf sample employed here
that there is indeed a break in the Tully-Fisher relation for faint
field\footnote{The $K'$-band data of Pierini \& Tuffs (1999\markcite{PT})
shows a steep slope with no break down to $V_c \approx 60\kms$.
These are cluster galaxies, so this makes
sense if these objects are less gas rich than the field sample.} galaxies. 
For $V_c \lesssim 90\kms$, galaxies are underluminous for their rotation
velocity as predicted by the extrapolation of a linear fit to the bright
galaxy data. There is a great amount of scatter here too --- the
relation bends and flares.  There have been concerns that there might
be curvature in the Tully-Fisher
relation (e.g., Bothun \& Mould 1987\markcite{curve})
but the data in Figure 1(a) are probably better described by a broken power
law, if it makes sense to fit anything to the faint end at all.

A break in the Tully-Fisher relation would have important ramifications for its
application and interpretation.  However, many of these faint galaxies are
very gas rich.  So much so, in fact, that the gas outweighs the stars in most
of them for any reasonable
choice of stellar mass-to-light ratio (Schombert, McGaugh, \& Eder
2000\markcite{SME}).  Therefore, we examine in Figure 1(b) the effects of
including the gas mass in the ordinate by plotting the total observed
baryonic disk mass, $\mass_d = \mass_* + \mass_{gas}$.  This has the remarkable
effect of restoring a single linear relation over the entire span of the 
observations.

It appears that the fundamental relation underpinning the Tully-Fisher
relation is one between rotation velocity and total baryonic disk mass. 
This relation has the form
\begin{equation}
\mass_d = {\cal A} V_c^b.
\end{equation}
An unweighted fit to the red ($I$, $H$, and $K'$-band) data gives
$\log{\cal A} = 1.57 \pm 0.25$
and $b = 3.98 \pm 0.12$.  The precise value of the normalization would of
course change if we assumed a different distance scale or different
stellar mass-to-light ratios.  The slope is indistinguishable from $b = 4$. 
If we fix the slope to this value, the normalization is
${\cal A} \approx 35\, (\Upsilon_*^{K'}/0.8)\,
h_{75}^{-2}\;\mass_{\sun}\;{\rm km}^{-4}\;{\rm s}^4$.

\section{Implications}

The basic result seen in Figure 1(b) falls directly out of
the observations.  All we have done is assume a plausible
mass-to-light ratio
for the stars, added in the gas mass, and plotted the data.
This simple result has a number of interesting implications.  

First, there is an apparently universal relation between baryonic mass and
rotation velocity, with a single normalization.   While this relation
specifically applies to our sample of late type spiral galaxies, it
seems plausible that it might also apply to early type spirals, provided
appropriate consideration is given to the bulge component, which
might require a different $\Upsilon_*$, and to any other baryonic
components which might be significant (like molecular gas).

The logarithmic slope of the relation is indistinguishable from 4.  
While this
slope is often attributed to the virial theorem, it is possible to derive
other slopes as well depending on the assumptions one makes
(Mo et al.\ 1998\markcite{MMW}).   Current cold dark matter
models predict a slope of 3 (Mo et al.\ 1998\markcite{MMW}; Steinmetz \&
Navarro 1999\markcite{SN}) which is excluded at $8 \sigma$.
Significant tweaking is required to obtain the observed slope.
Feedback from supernovae is often invoked in this context 
(van den Bosch 1999\markcite{vdB}),
but it is not obvious that the modest amount of feedback required by the
Tully-Fisher relation is consistent with the large amount needed to
explain the luminosity function (Lobo \&  Guiderdoni 1999\markcite{LB}).  
The correct slope and normalization is predicted by one
alternative to cold dark matter (Milgrom 1983\markcite{MOND}).
In this alternative there is no dark matter --- all of the mass is baryonic.

Whatever mechanism sets the observed relation is intimately 
connected to the
observed baryonic mass.  The interpretation of the standard 
luminosity-linewidth
relation has long supposed that the stellar mass-to-light ratios of
galaxies are a nearly uniform.  Indeed, the error budget allowed by
the modest amount of intrinsic scatter observed in the $K'$-band
is easily consumed by variations
in the star formation history (Verheijen 1997\markcite{Marc}).  There is
little room left for variation in the IMF, or cosmic scatter in 
the underlying mass-rotation velocity relation.

We have now addressed another piece of this puzzle.  In addition to the
near constancy of $\Upsilon_*$, we have explicitly corrected for the
stellar fraction $f_*$.  Equation (1) now reduces to
\begin{equation}
\mass_d = f_d f_b \mass_{tot}.
\end{equation}
The presumed mass-rotation velocity relation can now show through in
the observations provided both $f_b$ and $f_d$ are universal constants.
The baryon fraction of the universe is constant by definition.  But
it is less obvious that the fraction of baryons which reside in the disk
should be the same for all spirals.  Indeed, it is frequently suggested
(e.g., Navaroro, Eke, \& Frenk 1996\markcite{Eke}) that the sort of 
faint dwarfs studied here are likely to lose a significant
portion of their baryons.  This idea is blatantly at odds with the data,
as the product $f_d f_b$ would no longer\footnote{One could contemplate 
a variable $f_d$ provided that it was a {\it very\/} finely tuned 
(zero scatter) function of circular velocity.  For example,
$f_d \propto V_c$ would recover the slope predicted by CDM.} be constant.

It seems to us implausible that $f_d$ could be some arbitrary yet
universal fraction.  While it is easy to imagine mechanisms which
might prevent some of the baryons from cooling to join the disk,
it is difficult to contemplate any which do so with the required precision.
There is very little room in the budget for the intrinsic scatter for 
any scatter in $f_d$.  Let us call the mass in baryons
not already accounted for in the disk mass
$\mass_{other}$.  The disk fraction is then
\begin{equation}
f_d = \frac{\mass_* + \mass_{gas}}{\mass_* + \mass_{gas} + \mass_{other}}.
\end{equation}
If this other form of baryonic mass is significant ($\mass_{other} \sim
\mass_*$), then $f_d < 1$, but there should be a lot of
scatter in $f_d$ unless some magical mechanism strictly regulates the
ratio $\mass_{other}/(\mass_* + \mass_{gas})$.  This unlikely situation
occurs naturally only if 
$\mass_{other} \ll \mass_* + \mass_{gas}$, so $f_d \rightarrow 1$.
The modest intrinsic scatter in the Baryonic Tully-Fisher relation
therefore suggests that the luminous mass in stars and gas
represents nearly all the baryons associated with an individual galaxy
and its halo, arguing against a significant mass of dark baryons in
these systems.

\section{Conclusions}

We have explored the Tully-Fisher relation over five decades in 
luminous mass.  This is a considerable increase in dynamic range 
over previous studies.  We find clear evidence for a break in the 
optical Tully-Fisher relation around $V_c \approx
90\kms$.  Galaxies with rotation velocities less than this are
underluminous relative to the extrapolation of the fit to more 
rapidly spinning galaxies.  However, 
these faint galaxies are very gas rich.  Considering both stellar 
and gas mass restores a linear relation over the entire observed range.

These observations strongly suggest that the Tully-Fisher relation is
fundamentally a relation between rotation velocity and total baryonic
disk mass. 
This relation has the form
\begin{displaymath}
\mass_d = {\cal A} V_c^4
\end{displaymath}
with ${\cal A} \approx 35\, (\Upsilon_*^{K'}/0.8)\,
h_{75}^{-2}\;\mass_{\sun}\;{\rm km}^{-4}\;{\rm s}^4$.
The well known optical Tully-Fisher relation is an approximation to this
more fundamental relation in the limit of galaxies dominated by stars.

The existence of the Baryonic Tully-Fisher relation has a number of
important implications.  That it works means that
stars in spiral galaxies have mass-to-light ratios which are reasonable
for composite stellar populations.  The modest amount of scatter
indicates that the IMF must be nearly universal in order to
yield such uniform mass-to-light ratios.  
Only corrections for gas content are necessary to obtain the Baryonic
Tully-Fisher relation.  The data do not allow much room
for any further significant baryonic mass components.
Any component of dark baryons which does exist must do so in strict
proportionality to the observed baryons, with effectively zero scatter.
This unlikely situation argues against a significant mass in dark baryons
in any form (be it very cold molecular gas in the disk, very hot ionized 
gas in the halo, or baryonic MACHOs).  Any model which supposes a large 
mass of such baryons must explain why it contributes 
so little to the scatter in the Baryonic Tully-Fisher relation.

The results presented here make sense in terms of
a simple interpretation of the Tully-Fisher relation in which the
mass of observed baryons is directly proportional to the total mass which
in turn scales with the observed rotation velocity.  This
potentially includes the case where the mass observed in baryons {\it
is\/} the total mass (Milgrom 1983\markcite{MOND}).  Matching
these observations is a substantial challenge for modern structure
formation theories based on cold dark matter.  These predict a slope 
which is too shallow (3 rather than 4, different by $8\sigma$), and
fail to anticipate that effectively all the baryons associated with a halo
have cooled into the disk.

\acknowledgements We thank Ken Freeman for relevant conversations, and
the referee, H.-W. Rix, for many insightful comments.
The work of SSM is supported in part by NSF grant AST 99-01663.

\clearpage
{\tt
\noindent ADDENDUM: \\
The text above is rigorously identical to that accepted by {\it ApJ Letters}.
Two points were lost in the effort to meet the {\it Letters'\/} page limit. \\
{\bf 1: Maximum Disk Mass-to-Light Ratios:} We have assumed a constant
$\Upsilon_*$ which is plausible for a composite stellar population.  Presumably,
there is scatter about this value which is reflected in the non-observational
component of the scatter in Figure 1(b).  If instead we adopted the
stellar mass-to-light ratio suggested by maximum disk fits to rotation
curves, the resultant relation would have a much larger scatter.
The reason for this is that $\Upsilon_*^{max}$ increases
systematically with decreasing disk central surface brightness
(de Blok \& McGaugh 1997, MNRAS, 290, 533; Swaters, Madore, \& Trewhella 2000,
ApJ, 531, L107), yet there is a range of surface brightnesses at a given
luminosity.  For example, Swaters {\it et al.} discuss a number of galaxies
with $M_B \approx -18$ for which $\Upsilon_*^{max}$ ranges from 1.5 to 17.
Obviously, this factor would cause a large difference in the computed mass
at a given circular velocity, and inflate
the scatter in the Baryonic Tully-Fisher Relation.  This argues against
maximal disks in low surface brightness galaxies, though these remain
plausible for high surface brightness galaxies. \\
{\bf 2: Deviant Galaxies:} Recently, O'Neil, Bothun, \& Schombert 
(1999, AJ, 119, 136) have pointed out a population of red low surface
brightness galaxies.  Four of these galaxies deviate significantly from
the Baryonic Tully-Fisher relation, falling roughly an order of magnitude
below it (having
too little mass for their circular velocity).  Given their red colors,
these galaxies might have high $\Upsilon_*$, or could
contain some other component
of baryonic mass.  This point is tentative, so here we note only that the
Baryonic Tully-Fisher relation is a good diagnostic for identifying unusual
galaxies.  This also applies to galaxies which might have very low
$\Upsilon_*$ as a result of a young stellar population (e.g.,
Meurer {\it et al.} 1996, AJ, 111, 1551).
}

\vspace{2in}
\figcaption[f1.eps]{The Tully-Fisher relation plotted as a) stellar mass and
b) baryonic disk mass against rotation velocity.  Square symbols represent
galaxies where the circular velocity is estimated from the linewidth
by $V_c = \onehalf W_{20}$, while circles have $V_c = V_{flat}$ from
resolved rotation curves.  Data employed include the
$H$-band data of Bothun et al.\ (1985; red),
the $K'$-band data of Verheijen (1997; black), and
the $I$-band data of Pildis et al.\ (1997) with velocities
as reported by Eder \& Schombert (1999; green).
Also shown are the $B$-band data of McGaugh \& de Blok (1998; light blue),
and of Matthews et al.\ (1998; dark blue).
The stellar mass is computed from the
luminosity assuming a constant mass-to-light ratio: 
$\mass_* = \Upsilon_* L$, so (a) is
directly analogous to the usual luminosity-linewidth diagram.  
We assume mass-to-light ratios for the stellar populations
of late type galaxies of $\Upsilon_*^B = 1.4$, $\Upsilon_*^I = 1.7$,
$\Upsilon_*^H = 1.0$, and $\Upsilon_*^{K'} = 0.8 \;
\mass_{\sun}/L_{\sun}$ (see text).
In (b), we plot the total baryonic disk mass
$\mass_d = \mass_* + \mass_{gas}$ with $\mass_{gas} = 1.4 \mass_{HI}$. 
In (a), a clear break is apparent.  Galaxies with 
$V_c \lesssim 90\kms$ fall systematically below
the Tully-Fisher relation defined by brighter galaxies.  
In (b), the deficit
in mass apparent in (a) has been restored by including the gas mass.
The solid line is an unweighted fit to the red band data in (b)
with a correlation coefficient of 0.92 and a slope indistinguishable from 4.
\label{f1}}

\clearpage
\begin{figure}
\plotone[99 34 577 700]{Fig1.ps}
\end{figure}

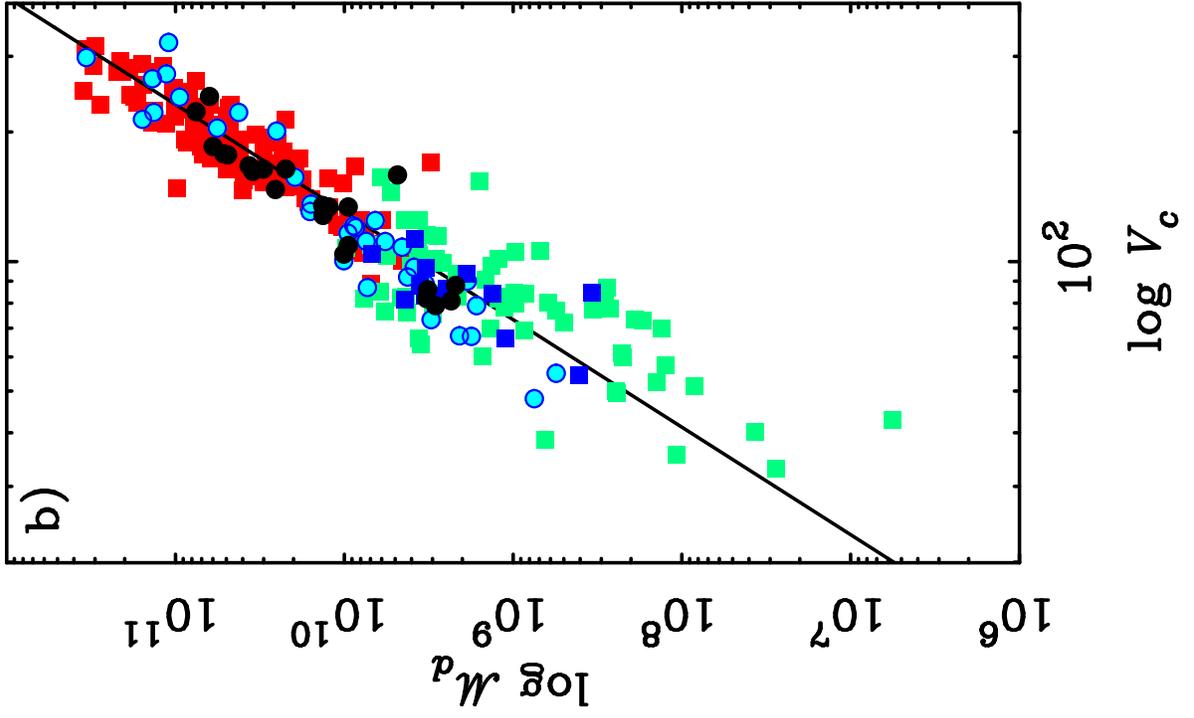
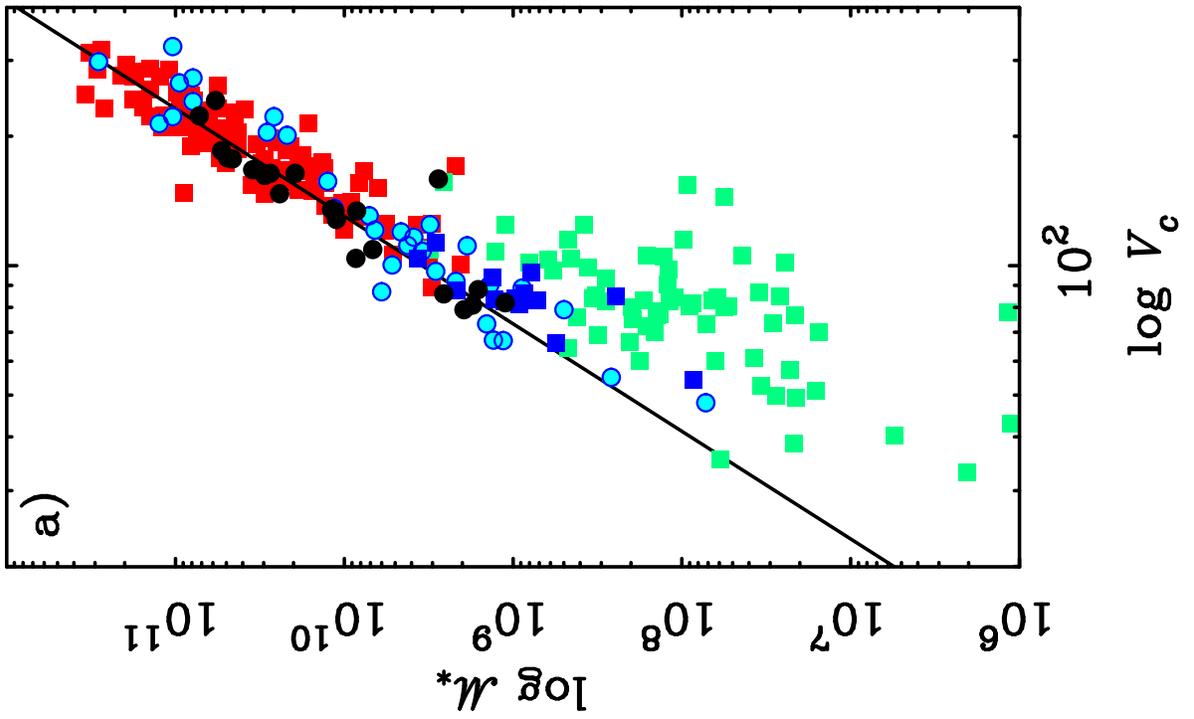
\end{document}